\pdfoutput=1

\documentclass[11pt]{article}

\usepackage[]{acl}

\usepackage{times}
\usepackage{latexsym}

\usepackage[T1]{fontenc}

\usepackage[utf8]{inputenc}

\usepackage{microtype}

\usepackage{graphicx}

\usepackage[utf8]{inputenc}
\usepackage{color,soul}
\setul{0.45ex}{0.25ex}
\usepackage{graphicx}
\usepackage{subcaption}
\usepackage{amsmath}
\usepackage{booktabs}
\usepackage{multirow, makecell}
\usepackage{tikz}
\usepackage{graphicx}
\usepackage{array}
\usepackage{pgfplots}
\usepackage{algorithm}
\usepackage{textcomp}
\usepackage{times}
\usepackage{latexsym}
\usepackage{amsmath}
\usepackage{amsfonts}
\usepackage[noend]{algpseudocode}
\usepackage{lipsum}
\usepackage{multirow}
\usepackage{framed}
\usepackage{graphicx}
\usepackage{pifont}
\usepackage{longtable}
\usepackage{tikz}
\usepackage{booktabs}
\usepackage[all]{nowidow}
\usepackage{amssymb}
\usepackage{enumitem}
\usepackage{color,soul}
\usepackage{bm}
\usepackage{arydshln}
\usepackage{xspace}
\usepackage{siunitx}
\usepackage{booktabs}
\usepackage{booktabs}
\usepackage{xspace}
\usepackage{adjustbox}

\usepackage{amsmath}

\usepackage{xcolor}
\definecolor{mypink1}{rgb}{0.858, 0.188, 0.478}
\definecolor{darkred}{rgb}{0.74,0.03,0}
\definecolor{mustardyellow}{rgb}{0.88,0.67,0.01}
\definecolor{navy}{rgb}{0,0,0.5}
\definecolor{darkcyan}{rgb}{0,0.54,0.54}

\newif\ifcomments
\commentstrue
\ifcomments
    \providecommand{\zhouhang}[2][]{{\protect\color{darkcyan}{[Zhouhang:\textbf{#1} #2]}}}

\else
    \providecommand{\zhouhang}[2][]{}
\fi

\usepackage{cleveref}

\title{On Faithfulness and Coherence of Language Explanations\\for Recommendation Systems}

\def\authorspace{\hspace{4mm}}

\author{
    Zhouhang Xie\authorspace{}
    Julian McAuley\authorspace{}
    Bodhisattwa Prasad Majumder\authorspace{}
        \\
        University of California, San Diego \\
        \{\href{mailto:zhx022@ucsd.edu}{\texttt{zhx022}}, \href{mailto:jmcauley@eng.ucsd.edu}{\texttt{jmcauley}}, 
        \href{mailto:bmajumde@eng.ucsd.edu}{\texttt{bmajumde}}\}\texttt{@ucsd.edu}
        }

\begin{document}
\maketitle
\begin{abstract}
Reviews contain rich information about product characteristics and user interests and thus are commonly used to boost recommender system performance.
Specifically, previous work show that jointly learning to perform review generation improves rating prediction performance.
Meanwhile, these model-produced reviews serve as recommendation explanations, providing the user with insights on predicted ratings.
However, while existing models could generate fluent, human-like reviews, it is unclear to what degree the reviews fully uncover the rationale behind the jointly predicted rating.
In this work, we perform a series of evaluations that probes state-of-the-art models and their review generation component.
We show that the generated explanations are brittle and need further evaluation before being taken as literal rationales for the estimated ratings.
\end{abstract}

\section{Introduction}

Product reviews capture rich information about user preferences and thus improve recommender system performance~\cite{McAuley2012LearningAA, McAuley2013HiddenFA, Zheng2017JointDM, Tay2018MultiPointerCN, Chen2018NeuralAR, Pugoy2020BERTBasedNC, pugoy-kao-2021-unsupervised}. 
Meanwhile, advancements in text generation enable generating realistic synthetic reviews conditioning on user and item identifiers, as well as additional features such as historical reviews~\cite{li-tuzhilin-2019-towards}, product metadata~\cite{ni-mcauley-2018-personalized, dong-etal-2017-learning-generate}, knowledge graph embedding~\cite{Li2021KnowledgebasedRG}, and sometimes the rating itself~\cite{Chen2021AspectLevelSR}.
Recently, there has been increasing interest in coupling rating estimation and review generation, treating generated reviews as \textit{explanations} for model recommendations~\cite{ni-etal-2017-estimating, Sun2020DualLF,li_2020_generate_neural_template, li-etal-2021-personalized, Hada2021ReXPlugER}. 

\begin{figure}[tb]
\centering
\includegraphics[width=0.85\columnwidth]{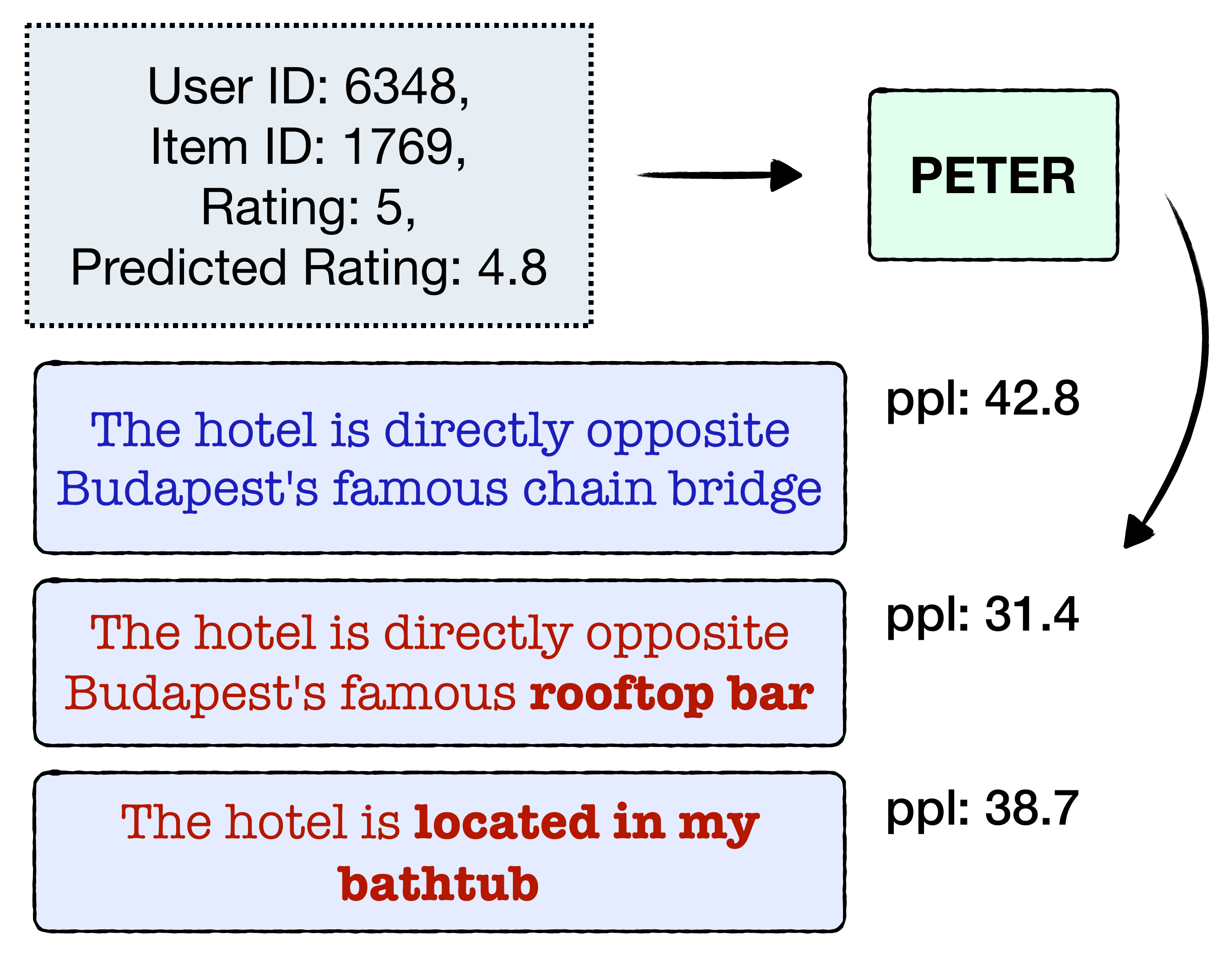}
\caption{PETER, a state-of-the-art model assigns lower perplexity to factually incoherent reviews.}
\vskip -3mm
\label{fig:broken_peter}
\end{figure}

In the current literature, the quality of the generated explanations are usually measured by perplexity and overlapping-based metrics such as Distinct-N~\cite{li-etal-2016-diversity}, Rouge score~\cite{lin-2004-rouge}, and BLEU score~\cite{papineni-etal-2002-bleu} with respect to the ground truth reviews. 
However, while these evaluations measure fluency and word-overlapping, they do not warrant the the generated reviews' quality as explanations. 

Specifically, overlapping metrics overlook two core aspect of natural language explanations (NLEs): (1) \textit{faithfulness}, how truthfully do the generated explanations reflect the decision process for the models rating prediction, and (2) \textit{semantic coherence}, how well the model capture the users' true interest towards the product.
To highlight the potential issue associated with current evaluation, consider the review text for a restaurant "\textit{I love this hotel because it has great service}" with a rating of 5, where the explanation generated is "\textit{I love this hotel because it has great cookies}" with the correct predicted rating. 
The generated explanation deviates from the ground truth sentence by only one word, yet completely changes the rationale for the rating. 
However, the currently widely used automatic metrics will still assign a high score to the generated review.
Further, there is no guarantee that even \textit{cookie} is truly accountable for the predicted rating.

\begin{figure*}[tb]
\centering
\includegraphics[width=0.9\textwidth]{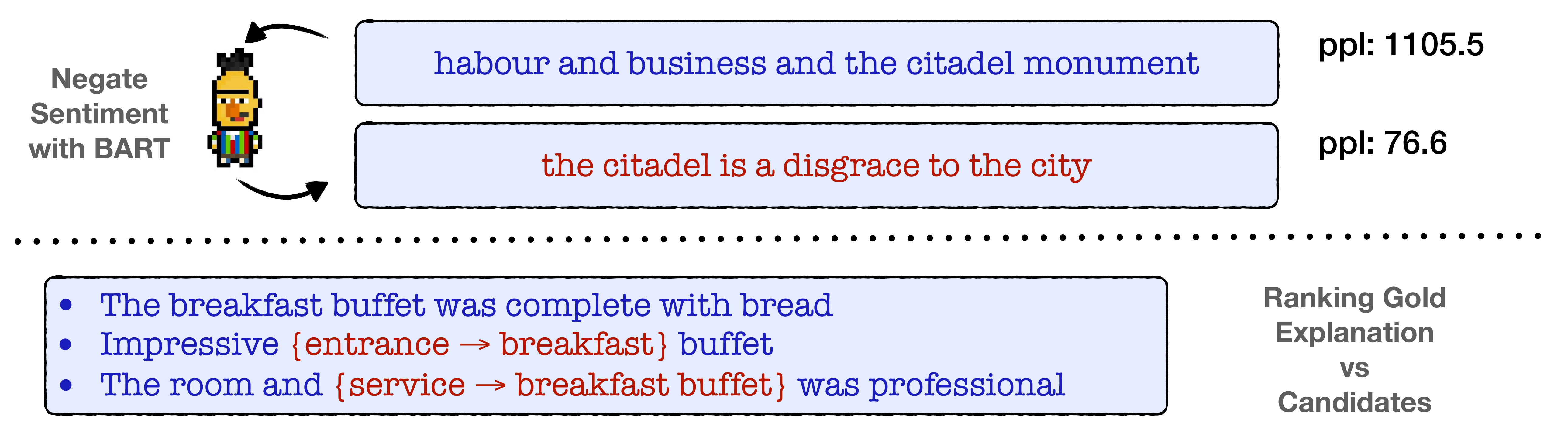}
\caption{\textbf{Up:} evaluating adversarial invariance ratio. \textbf{Down:} robustness of review generation by ranking against alternative explanations. Text snippet in \textcolor{darkred}{red} denotes perturbed content. See \Cref{para:air} for details.}
\vskip -3mm
\label{fig:main_evaluation_schema}
\end{figure*}

To address these discrepancies, we argue that NLEs for recommendation systems should be evaluated as explanations, similar to NLEs in NLP tasks.
\textbf{In this work}, we probe review-as-explanation models in explainable recommendation literature.
Our results show a concerning trend that current models struggle to produce reviews that are semantically coherent with the ground truth reviews, and are inconsistent with the explanation they produce.
We encourage researchers and practitioners take beyond-overlapping evaluations into account when training review generation models for explanations. 
Better evaluations could lead to deeper understanding of capabilities of these generated rationales, and foster more trustworthy explainable recommendation systems.

\section{Problem Definition and Models}

This section is organized as follows: we first cover the task of joint review-rating generation, then introduce the models used in our experiments.

\subsection{Problem Setup}

Given a user $u$ and an item $i$, the task of joint review-rating generation aims at predicting an associated rating $\hat{r}$ as well as a natural language explanation $\hat{e}$~\footnote{which is commonly the associated review}. 
During training, the model jointly minimizes the negative log likelihood (NLL) of ground truth reviews in the corpus, as well as the mean squared error (MSE) of their associated rating.

\subsection{Models}

We compare four recent models in the literature that covers a variety of commonly used architectures in natural language generation: Att2Seq~\cite{dong-etal-2017-learning-generate}, NRT~\cite{Li2017NeuralRR}, PEPLER~\cite{2022-PEPLER}, and PETER~\cite{li-etal-2021-personalized}. Among the models, Att2Seq and NRT are based on Long Short-Term Memory (LSTM)~\cite{Hochreiter1997LongSM}, while PEPLER combines a pre-trained GPT-2 model~\cite{Radford2019LanguageMA} with prompt tuning. Finally, PETER adopts non-auto-regressive transformer architecture. 

Meanwhile, since~\citeauthor{li-etal-2021-personalized} noted incorporating specific content words significantly improves generation quality, we follow the original paper and condition the model on a content word, denoted by PETER$_{cond}$. 
Note that following the previous implementation, the aspect word in the dataset is extracted from \textit{ground truth} review, giving PETER$_{cond}$ an unfair advantage. We thus use PETER$_{cond}$ as an upper-bound baseline.

\subsection{Datasets}

 We conduct our experiments on Yelp\footnote{https://www.yelp.com/dataset/challenge} (Y.), TripAdvisor\footnote{https://www.tripadvisor.com
} (T.), and Movies and TV category from Amazon dataset~\cite{He2016UpsAD} (M.).
These are standard dataset commonly used to benchmark joint review-rating estimation models~\cite{li_2020_generate_neural_template}.

\section{Evaluating Faithfulness and Semantic Coherence}

We generate 10,000 explanations for each model on each dataset, and perform a set of evaluations as described in this section.

\subsection{Faithfulness}

When reviews are treated as natural langauage explanations, joint review-rating prediction models could be categorized into self-rationalizing models. 
\citeauthor{jacovi-goldberg-2020-towards} argues that the quality of NLE~\footnote{NLR in \citeauthor{jacovi-goldberg-2020-towards}'s work} should be evaluated by both their plausibility, how convincing the explanations are to humans, and faithfulness, how truthful they reflect the models' decision process.
We focus on model faithfulness in this section. 

By definition, a faithful explanation will truthfully represent the decision process of a model. 
However, directly measuring faithfulness is infeasible due to the black-box nature of deep neural networks. 
We instead design a set of proxy tasks that test \textit{unfaithful} behavior of joint review-rating estimation models.

\paragraph{Adversarial In-variance Ratio (AIR).}\label{para:air} 
Since the explanation generation by the model is representative of the model's belief of the reasons behind the rating prediction, we argue that such belief must be robust to sentiment perturbations. 
In other words, assume a model generates a sequence $\hat{e}_{u,i}$ as an explanation, the sentiment-negated counter explanation $\neg \hat{e}_{u,i}$ should not receive a higher likelihood (lower perplexity) than the original review. 
Illustration of selected sub-experiments are as shown in \Cref{fig:main_evaluation_schema}.

Concretely, we take 4 or 5-star (positive sentiment) ratings from the test set and rewrite their sentiment to negative using a pretrained BART model\footnote{\url{dapang/yelp_pos2neg_lm_bart_large} from huggingface.}, and let the target model rank the ground truth and rewritten review with perplexity. 
We mark the models' decision as flipped if it assigns lower perplexity to the rewritten review with the negated sentiment.
In this case, the model's explanation is thus unfaithful.
Note that this means a random baseline would achieve 50 percent in AIR.

\paragraph{Mean Reciprocal Rank against Alternative Explanations (MRR-AE).} \label{para:tlae}
As pointed out in~\citeauthor{jacovi-goldberg-2020-towards}, a model is unfaithful if it provides a different interpretation for the same decision by the same model.
That is, the model should be able to differentiate its generated review from other candidate explanations. 
Following this intuition, we argue that the model should have the ability to pick out its generated review from other reviews, such as random reviews drawn from the dataset or adversarially constructed ones, as shown in \Cref{fig:main_evaluation_schema}.

To measure this, we sample 100 reviews randomly from the test dataset for each gold review, and replace the aspect in the sampled sentences with the aspect covered by the ground truth. 
We then let the target model rank the 100 
sentences along with the gold review with perplexity score and measure its performance with mean reciprocal rank (MRR). 
The random baseline for MRR-AE is thus around 5 percent.

\paragraph{Text-label Agreement Error (TLAE).} 
As faithful explanations, the generated reviews should strongly correlate with predicted ratings. 
To measure this, we train a BERT~\cite{vaswani2017attention} based auxiliary rating regressor based on \textit{only} user reviews on the training set of the models being evaluated. 
At test time, we measure the Mean Squared Error of the auxiliary predictor on \textit{generated} reviews and regressor-predicted ratings.

\begin{table*}[h]
\small
\setlength\tabcolsep{4pt}
\center
\resizebox{\textwidth}{!}{%
\begin{tabular}{l cc cc cc cc cc cc c cc cc cc cc cc cc cc cc cc c cc cc cc@{}}
\toprule
\multirow{2}{*}{}
  & \multicolumn{9}{c}{\textbf{Faithfulness}}
  &
  & \multicolumn{9}{c}{\textbf{Semantic Coherence}} 
  &
  & \multicolumn{3}{c}{\textbf{Rec.}}
  \\
\cmidrule(lr){2-10}
\cmidrule(lr){11-20}
\cmidrule(lr){22-24}
  \textbf{Metric} 
  & 
  \multicolumn{3}{c}{\textbf{AIR}$\uparrow$} & \multicolumn{3}{c}{\textbf{MRR-AE}$\uparrow$} & \multicolumn{3}{c}{\textbf{TLAE}$\downarrow$} &
  &
  \multicolumn{3}{c}{\textbf{Entail}$\uparrow$} & 
  \multicolumn{3}{c}{\textbf{BERTS.}$\uparrow$} & \multicolumn{3}{c}{\textbf{ BARTS.}$\downarrow$} &  
  & 
 \multicolumn{3}{c}{\textbf{RMSE}$\downarrow$} &
 \\
\cmidrule(lr){2-4}
\cmidrule(lr){5-7}
\cmidrule(lr){8-10}
\cmidrule(lr){12-14}
\cmidrule(lr){15-17}
\cmidrule(lr){18-20}
\cmidrule(lr){22-24}
  \textbf{Model}
  & 
  \multicolumn{1}{c}{\textbf{M.}} &
  \multicolumn{1}{c}{\textbf{T.}} & \multicolumn{1}{c}{\textbf{Y.}} &
  \multicolumn{1}{c}{\textbf{M.}} &
  \multicolumn{1}{c}{\textbf{T.}} & \multicolumn{1}{c}{\textbf{Y.}} &  \multicolumn{1}{c}{\textbf{M.}} &
  \multicolumn{1}{c}{\textbf{T.}} & \multicolumn{1}{c}{\textbf{Y.}} & 
  &
  \multicolumn{1}{c}{\textbf{M.}} &
  \multicolumn{1}{c}{\textbf{T.}} & \multicolumn{1}{c}{\textbf{Y.}} & \multicolumn{1}{c}{\textbf{M.}} &
  \multicolumn{1}{c}{\textbf{T.}} & \multicolumn{1}{c}{\textbf{Y.}} & \multicolumn{1}{c}{\textbf{M.}} &
  \multicolumn{1}{c}{\textbf{T.}} & \multicolumn{1}{c}{\textbf{Y.}} & 
  & 
 \multicolumn{1}{c}{\textbf{M.}} &
  \multicolumn{1}{c}{\textbf{T.}} & \multicolumn{1}{c}{\textbf{Y.}} &
 \\
\midrule
Att2Seq
               & 14.6 & 43.5 & 47.9
               & 23.5 & 19.2 & 23.5 
               & n/a & n/a & n/a
               &
               & 6.6 & 2.7 & 5.6
               & 0.08 & 0.16 & 0.10
               & 5.95 & 5.97 & 5.97
               &
               & n/a & n/a & n/a
\\
\midrule
NRT
               & 55.8 & 43.0 & \underline{\textbf{55.8}}
               & 15.1 & 18.2 & 22.6
               & \textbf{1.39} & 0.84 & \textbf{1.18} 
               &
               & 3.2 & 1.5 & 1.4
               & -0.21 & -0.15 & -0.19
               & 6.54 & 6.72 & 6.6
               &
               & \textbf{0.95} & \underline{\textbf{0.79}} &  \underline{\textbf{1.01}}
\\
\midrule
PETER
               & 60.3 & 47.3 & 49.4
               & 18.1 & \underline{\textbf{22.0}} & 26.3
               & \textbf{1.39} & 0.84 & \textbf{1.18}  
               &
               & 8.2 & 3.5 & 7.9  
               & 0.11 & 0.17 & 0.12     
               & 5.95 & 5.96 & 5.88
               &
               & \textbf{0.95} & 0.81 & \underline{\textbf{1.01}}
\\
\midrule
PEPLER
               & \underline{\textbf{70.0}} & \underline{\textbf{63.3}} & 21.0
               & 16.6 & 16.3 & 7.3
               & \textbf{1.39} & 0.84 & \textbf{1.18} 
               &
               & 4.4 & 3.4 & 4.6
               & 0.13 & 0.19 & 0.23 
               & 5.95 & 5.93 & 6.99
               &
               & 1.25 & 1.71 & 1.69
\\
\midrule
PETER$_{cond}$
               & 19.1 & 52.0 & 55.7
               & \textbf{27.9} & 20.2 & \textbf{27.9}
               & \textbf{1.39} & \textbf{0.76} & \textbf{1.18} 
               &
               & \textbf{27.7} & \textbf{24.2} & \textbf{24.3}
               & \textbf{0.18} & \textbf{0.30} & \textbf{0.25}  
               & \textbf{5.47} & \textbf{5.93} & \textbf{5.17}
               &
               & \textbf{0.95} & 1.81 & 1.02
\\
    \bottomrule
    \end{tabular}
}
    \caption{Evaluation results on the datasets. \textbf{Bold text} denotes best performance (except PETER$_{cond}$ mdel), \underline{\textbf{bold and underlined}} text means the performance exceeded PETER$_{cond}$ model, which has access to ground truth aspect covered in review. M., T. and Y. denotes Amazon Movies and TV, TripAdvisor, and Yelp dataset, respectively.}
\vskip -3mm
\label{tab:main_evaluations}
\end{table*}

\subsection{Semantic Coherence}

Traditional evaluation metrics use in the literature focuses on word-overlapping, and thus would be insensitive to mismatched content words.
To address this issue, we argue that generated explanations should be evaluated by its semantic coherence. 
Concretely, we adopt two recent, state-of-the-art semantic evaluation metrics: BERTScore~\cite{bert-score} and BARTScore~\cite{yuan2021bartscore}\footnote{BERTScore is cosine similarity-based (larger means better) and BARTScore is NLL based (smaller means better).}. 
Further, we use a pre-trained entailment model to check whether the generated content entails the ground truth review. 
We report the percentage of entailment (Entail), where a good model should have high ratio.

\section{Empirical Results}

\paragraph{Faithfulness.} Our main evaluation results are as shown in table~\ref{tab:main_evaluations}. 
While model-generated explanation generally matches the predicted rating (TLAE), most models have near random performance against sentiment perturbations (AIR).
Meanwhile, although PEPLER is the most robust to sentiment perturbation, it is not as competitive as other models in terms of recommendation performance (RMSE).
This illustrates the potential risk of powerful language models giving a false sense of explainability simply due to their strong language modeling ability.
In other words, the explanations are plausible but not faithful under~\cite{wiegreffe-etal-2021-measuring}'s framework.

\paragraph{Semantic Coherence.} From coherence evaluations, we could see the model generally struggle to capture the exact aspect that the user cares about, resulting in a low entailment ratio (Entail) compared to PETER$_{cond}$.
This can be corroborated by BERTScore and BARTscore, highlighting the importance of conducting semantic evaluations for explanation generation.

\section{Discussion and Analysis}

\paragraph{Non-robust correlation between generated review and estimated rating.}
\label{para:non_robust_corr}
Based on TLAE, we could observe that the generated review is indeed correlated to the predicted rating. 
However, the models being evaluated all demonstrate near-random AIR scores, showing such correlations are brittle, and the language model's belief is entangled with other rating irrelevant factors.

\paragraph{Weak correlation between generated item aspect and the item.}
\label{para:weak_aspect_corr}
From the MRR-AE score, we could see that models generally perform poorly in ranking the generated review against synthetic alternative explanations, where the description of a random item is used in place of the generated one. 
This behavior shows the model fails to establish robust connections between generated reviews and the corresponding items.

\paragraph{How much can reviews explain rating?}
While reviews boost recommender system performance, they cannot \textit{fully} explain the corresponding rating. 
In particular, human-written explanations are inherently limited in \textit{discovery} tasks, where a machine learning model needs to demonstrate beyond-human performance~\cite{tan-2022-diversity}.
As a result, human-written explanations are not the complete reason for inferring the label.
Consequently, maximizing the likelihood of explanations does not guarantee the best task performance~\cite{carton-etal-2022-learn}, which could be corroborated by findings from the recommender system community~\cite{DBLP:conf/sigir/SachdevaM20} and PEPLER model from our evaluation.

\section{Implications and Recommendations}

\paragraph{Calibrate user expectations.}
End users often trust that algorithm explanations are faithful~\cite{DBLP:conf/aaai/JinLH22}\footnote{\citeauthor{DBLP:conf/aaai/JinLH22} study was based on medical image} and hope they could receive reliable explanations~\cite{lakkaraju2022rethinking}, we recommend practitioners inform the users about the probablistic nature of generated reviews as explanations. For example, if a system generates an explanation related to ‘bbq pork ribs’, it would be more of an indication of user's interest in smokehouse cuisines rather than the dish itself.

\paragraph{Develop hybrid systems.}
While PETER$_{cond}$ acts as an ``unfair'' baseline in our experiment section, the model would be a great tool in a larger pipeline, where users actively provide feedback. 
Similarly, recent works starts to explore natural language as an \textit{interface} in pipeline systems for non-language based explanations~\cite{Slack2022TalkToModelUM}.
We encourage the community to consider conditional and pipeline systems in addition to end-to-end models.

\paragraph{Better evaluations.} 
We note that although evaluations generally depend on the use-cases of the model, and a powerful model does not necessarily need to satisfy faithfulness, plausibility, and semantic coherence simultaneously,
it is advisable to perform beyond-overlapping evaluations before assuming the literal validity of generated recommendation NLRs.

\subsection{Related Works}

\paragraph{Faithfulness of Natural Language Explanations.}
\citeauthor{jacovi-goldberg-2020-towards} argues that the quality of NLE from this class of models should be evaluated by %
their faithfulness, how truthful (and thus consistent) do they reflect the models' decision process.
Under this setup, our work's evaluation differs from existing evaluations in the literature in that we clearly distinguishes faithfulness from general language quality. \citeauthor{wiegreffe-etal-2021-measuring} approach this problem by measuring the connection between labels and explanations, yet their evaluation do not take the semantics of the generated explanation itself into account.

\paragraph{Analyzing Model Decision in NLP.}
Another related line of work is analysis of model decision boundaries. Common strategies usually involves adversarially probing the model, such as using counterfactual data~\cite{polyjuice:acl21}, constrast sets~\cite{gardner-etal-2020-evaluating} and semantically preserving modifications of sentence characteristics~\cite{ribeiro-etal-2018-semantically, ribeiro2020beyond, longpre-etal-2021-entity}. 
Our work deviates from prior works as we establish connections of adversarial evaluation directly with model faithfulness.

\section{Conclusions}

Joint review-rating prediction models could generate high-quality reviews while producing accurate rating estimations. 
However, it is unclear whether the generated reviews could be leveraged as precise recommendation rationales. 
We conduct a set of evaluation that benchmark faithfulness and semantic coherence of state-of-the-art models. 
We show more careful evaluations are needed before generated reviews could be taken as fully accountable explanations.

\bibliography{anthology,custom}
\bibliographystyle{acl_natbib}

\appendix

\end{document}